\documentclass[letterpaper,11pt]{article}
\usepackage{caption}
\pdfoutput=1 
\usepackage{xcolor}
\usepackage{enumerate}
\usepackage{framed}
\usepackage{mdframed}
\usepackage{amsmath}
\usepackage{amsfonts}
\usepackage{mathrsfs}
\usepackage{amssymb}
\usepackage[normalem]{ulem}
\usepackage{graphicx, rotating}
\usepackage{epsfig}
\usepackage{latexsym}
\usepackage{graphicx}
\usepackage{color}
\usepackage{soul}
\usepackage{amsmath,bm,amssymb}
\usepackage{cite}
\usepackage{slashed}
\usepackage{hyperref}
\usepackage{epstopdf}
\usepackage[font=footnotesize,labelfont=]{caption}
\AppendGraphicsExtensions{.tif}
\hypersetup{colorlinks=true, citecolor=bluscuro, linkcolor=black, urlcolor=bluscuro}
\definecolor{rossos}{cmyk}{0,1,1,0.55}
\definecolor{bluscuro}{rgb}{0.15, 0.2, .85}
\definecolor{bluchiaro}{cmyk}{1,.3,0.,0.1}

\newcommand{\be}{\begin{equation}}
\newcommand{\ee}{\end{equation}}
\newcommand{\bea}{\begin{eqnarray}}
\newcommand{\eea}{\end{eqnarray}}
\newcommand{\beas}{\begin{eqnarray*}}
\newcommand{\eeas}{\end{eqnarray*}}

\newcommand{\rd}{{\rm d}}

\begin{document}
\def\thefootnote{\fnsymbol{footnote}}

\begin{center}
\LARGE{\textbf{Implications of the Weak Gravity Conjecture for Tidal Love Numbers of Black Holes}} \\[0.5cm]
 
\large{Valerio De Luca\footnote{vdeluca@sas.upenn.edu}, Justin Khoury\footnote{jkhoury@sas.upenn.edu} and Sam S. C. Wong\footnote{scswong@sas.upenn.edu}}
\\[0.5cm]

\small{
\textit{Center for Particle Cosmology, Department of Physics and Astronomy, University of Pennsylvania,\\ Philadelphia, PA 19104}}

\vspace{.2cm}

\end{center}

\vspace{.6cm}

\hrule \vspace{0.2cm}
\centerline{\small{\bf Abstract}}
%\vspace{-0.2cm}
{\small\noindent 
The Weak Gravity Conjecture indicates that extremal black holes in the low energy effective field theory should be able to decay. This criterion gives rise to non-trivial constraints on the coefficients of higher-order derivative corrections to gravity. In this paper, we investigate the tidal deformability of neutral black holes due to higher-order derivative corrections. As a proof of concept, we consider a correction of cubic order in the Riemann curvature tensor. The tidal Love numbers of neutral black holes receive leading-order corrections from higher-order derivative terms, since black holes in pure General Relativity have vanishing tidal Love number. We conclude that the interplay between the tidal deformability of black holes and the Weak Gravity Conjecture provides useful information about the effective field theory. 
}

\vspace{0.3cm}
\noindent
\hrule
\def\thefootnote{\arabic{footnote}}
\setcounter{footnote}{0}

\section{Introduction} 
The recent discovery of gravitational waves has 
provided a powerful new tool to dig deeper into open questions regarding the interaction of fundamental theory and observation, in particular in probing gravity in its most extreme regime and in searching for signatures of new physics.
One of the most important prediction of General Relativity (GR) are black holes. Within this theory, the black hole mass, angular momentum, and electric charge uniquely define its entire multipolar structure as well as its quasinormal modes spectrum, whose observation can be used to probe the so-called no-hair theorems and to test GR~\cite{LIGOScientific:2021sio}.

The multipolar structure of a compact object may be modified under the presence of external fields, whose gravitational interaction can result into tidal deformations. Such effects are usually captured in terms of the tidal Love numbers (TLNs)~\cite{poisson_will_2014}. These strongly depend on the internal properties and structure of the deformed body. TLNs are also found to affect the dynamics of the inspiral of a binary system of compact objects and impact the consequent gravitational waves emission at the fifth post-Newtonian order~\cite{Flanagan:2007ix}. 

A powerful result of GR is that the TLNs of non-rotating and spinning black holes are precisely zero~\cite{Binnington:2009bb,Damour:2009vw,Damour:2009va,Pani:2015hfa,Pani:2015nua,Gurlebeck:2015xpa,Porto:2016zng,LeTiec:2020spy, Chia:2020yla,LeTiec:2020bos}. This result has generated an issue of “naturalness” in the gravitational theory~\cite{Porto:2016zng}, and it has been connected with special symmetries of the perturbation fields around black holes~\cite{Hui:2020xxx,Charalambous:2021kcz,Charalambous:2021mea,Hui:2022vbh,Charalambous:2022rre,Ivanov:2022qqt,Katagiri:2022vyz} or to relations between conformal field theories and black hole perturbations~\cite{Bonelli:2021uvf,Kehagias:2022ndy}.
This property is, however, broken in higher dimensions~\cite{Kol:2011vg,Cardoso:2019vof, Hui:2020xxx,Pereniguez:2021xcj} and especially in the context of modified gravity~\cite{Cardoso:2017cfl,Cardoso:2018ptl,Cvetic:2021vxa}. In this paper, as an example case, we will be interested in a theory with a $\left(R^{\mu\nu}_{\;\;\;\, \rho \sigma}\right)^3$ correction. Such an operator naturally appears when one includes six derivative terms in higher-derivative gravity~\cite{Metsaev:1986yb, Endlich:2017tqa, Ruhdorfer:2019qmk} and may also be generated at one-loop by integrating out massive fields with coupling given by~\cite{Goon:2019faz}
\be
\alpha = \frac{M_{\rm Pl}^2}{30240 (4\pi)^3} \sum \left( \frac{1}{m_s^2} - \frac{4}{m_f^2} + \frac{3}{m_v^2} \right),
\ee
in terms of the masses $m_s$, $m_f$, $m_v$ of a spin-0, spin-1/2  and spin-1 field, respectively.
We will argue that this operator represents the leading higher-order curvature correction that gives rise to non-vanishing TLNs for neutral black holes. 

Aside from the TLNs or quasinormal modes, black holes also provide essential information about quantum gravity. The strong gravity environment created by black holes is sensitive to corrections to GR. Especially small black holes are important as space-time curvature gets strong at the horizon. Hawking radiation is one example. Effective field theory (EFT) provides a convenient framework to systematically study higher-derivative corrections to GR. Most of the complicated microscopic degrees of freedom are effectively captured by effective operators at low energies.  

However, it is well known that not every bottom-up EFT of gravity can be consistently UV completed. Theories that do not meet this requirement are said to be living in the Swampland \cite{Vafa:2005ui}. Of many UV consistency conditions, the Weak Gravity Conjecture (WGC)~\cite{Arkani-Hamed:2006emk} is a well-studied criterion for consistent quantum gravity theories. It is conjectured that there must exist a state with charge to mass ratio (in proper units) larger than unity. There has been evidence for the WGC from the consideration of holography \cite{Nakayama:2015hga,Harlow:2015lma,Benjamin:2016fhe,Montero:2016tif}, black hole entropy \cite{Shiu:2016weq,Hebecker:2017uix,Cheung:2018cwt}, cosmic censorship \cite{Horowitz:2016ezu,Shiu:2016weq,Crisford:2017gsb,Yu:2018eqq}, dimensional reduction \cite{Brown:2015iha,Brown:2015lia,Heidenreich:2015nta,Heidenreich:2016aqi,Lee:2018urn} and importantly, IR consistency \cite{Cheung:2014ega,Andriolo:2018lvp,Hamada:2018dde,Bellazzini:2019xts,Arkani-Hamed:2021ajd}. The form of WGC we utilize deals with extremal black holes in the EFT \cite{Kats:2006xp, Cheung:2014ega, Cheung:2018cwt, Hamada:2018dde, Noumi:2022ybv}. It simply demands that black holes are able to decay and, as a consequence, the small extremal black holes are the states that satisfy the charge to mass ratio bound. 

In this paper we establish for the first time a connection between TLNs of black holes and UV consistencies of the theory, in particular related to the WGC. As a proof of concept, we focus on an admittedly tuned EFT containing only a $\left(R^{\mu\nu}_{\;\;\;\, \rho \sigma}\right)^3$ operator. This has the advantage of being the leading operator beyond GR affecting the TLNs of neutral black holes. We then show its impact on the black holes' tidal deformability and extremality relation.
Let us stress, however, that the argument is completely general and can be applied as well to lower-dimensional operators in the context of charged black holes. We compute the spectrum of extremal black holes in the theory of interest, and derive a bound on the coefficient of the operator. This constraint carries over to TLNs of black holes in the theory. Hence, measuring tidal deformations of black holes can potentially provide valuable information about the effective operators and particle spectrum of the UV theory.

The paper is organised as follows. In Section 2 we first extract the WGC constraint on coefficient of the higher-derivative operator~$\left(R^{\mu\nu}_{\;\;\;\,\rho \sigma}\right)^3$ by considering the spectrum of extremal black holes. Then, in Sections 3 and 4 we compute TLNs in the same theory for neutral black holes. We will conclude that the WGC imposes a type of positivity bound on the tidal deformability of black holes.

\section{Weak Gravity Conjecture constraints on extremal black holes}
In this Section we review the basic idea of the WGC and then apply the conjecture to derive constraints on the theory of interest. The original conjecture \cite{Arkani-Hamed:2006emk} states that a consistent theory of quantum gravity should contain a state of charge to mass ratio (in proper units) greater than unity. There are different aspects and different versions of the WGC. We will focus on the infrared consistency of low energy effective theories that stems from the WGC \cite{Kats:2006xp,Cheung:2014ega,Cheung:2018cwt,Hamada:2018dde,Noumi:2022ybv}, which requires that extremal black holes in the effective field theory be able to decay into sub-extremal black holes. 

Typically, in an effective field theory of gravity, the black hole geometry is corrected in the presence of higher-derivative operators. Corrections to the geometry coming from curvature invariants such as $R^{\mu\nu}_{\;\;\;\rho\sigma} R^{\rho\sigma}_{\;\;\; \mu\nu}$, are proportional to some negative power of the size of the black hole. Therefore higher-derivative corrections are more important as the size of the black hole decreases. This is also consistent with the intuition that smaller black holes are more sensitive to UV degrees on freedom. 

\begin{figure}[t]
    \centering
    \includegraphics[scale=0.6]{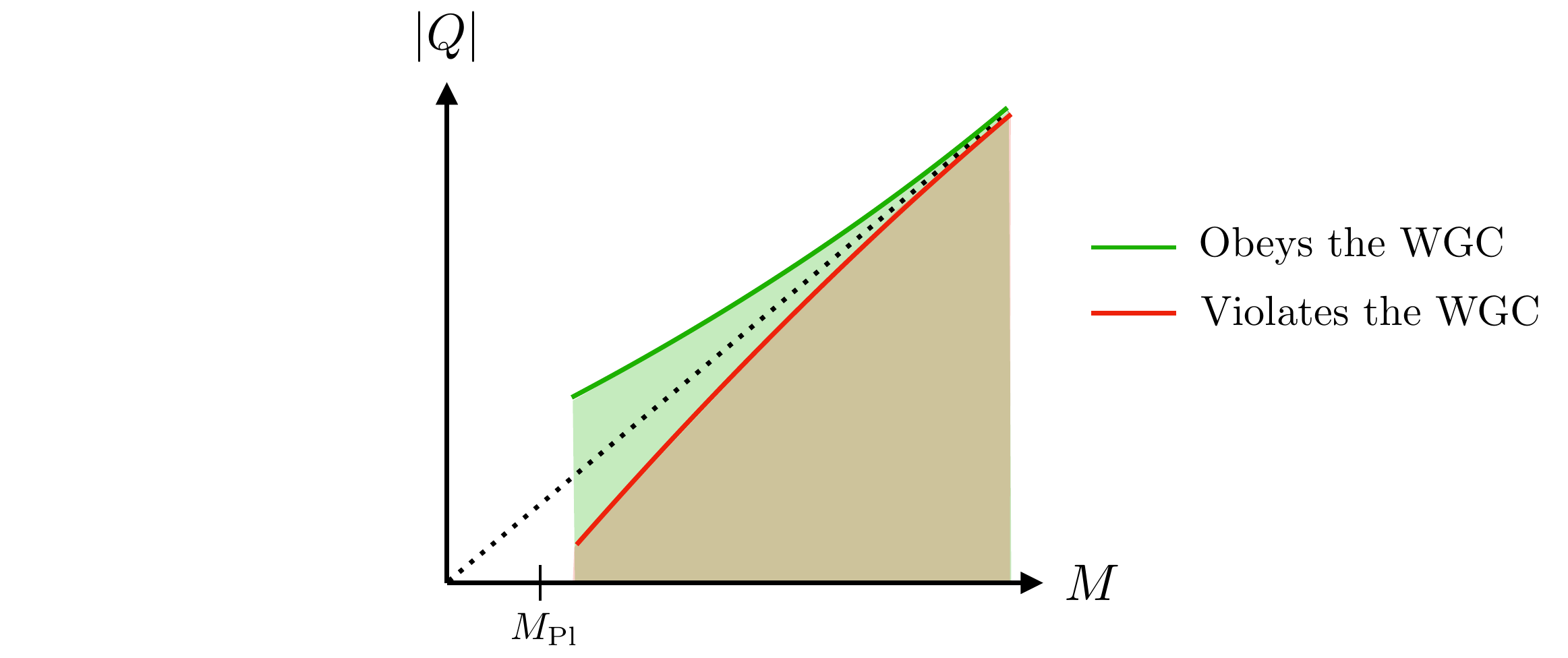}
    \caption{The curves are indicating the charge $|Q|$ and mass $M$ of extremal black holes in different effective theories. The dotted line is the GR extremality relation. Sub-extremal black holes are under the curves. Black hole configurations above the curves are forbidden as they contain naked singularities. The green curve represents a healthy effective field theory, since extremal black holes can decay into sub-extremal black holes. The red curve represents an effective field theory in which decay products of an extremal black hole must contain states above the curves, and therefore violates the weak gravity conjecture.}
    \label{fig:WGC}
\end{figure}

When it comes to extremal black holes, the charge to mass ratio $|Q|/M$ is corrected in a way that the decay of extremal black holes becomes a nontrivial story, as depicted in Fig.~\ref{fig:WGC}. If the corrected extremal relation $|Q|/M$ is below the GR relation, illustrated by the red curve in the figure, then any decay of an extremal black hole into macroscopic black holes must contain a super-extremal black hole with naked singularity. Therefore, this is forbidden by the WGC. On the contrary, if the corrected extremal relation lies above the GR relation (illustrated by the green curve), then extremal black holes can decay into sub-extremal black holes. Consequently, the WGC effectively imposes constraints on the effective operators. It turns out that this constraint is consistent with the relation between corrections to entropy and extremality derived from the thermodynamic point of view~\cite{Loges:2019jzs, Goon:2019faz, Cremonini:2019wdk,  Aalsma:2021qga}.

We now study extremal black holes in the specific theory of interest. 

\subsection{Charged black holes in $\left(R^{\mu\nu}_{\;\;\;\, \rho \sigma}\right)^3$ }
To demonstrate the idea that WGC constrains tidal deformations of black holes, we consider a simple theory, namely Einstein-Maxwell gravity corrected by a term cubic in Riemann:
\begin{align} \label{eqn:eft}
   S= -\int {\rm d}^4 x \sqrt{-g} \left[\frac{R}{2 \kappa^2} - \frac{1}{4} F_{\mu \nu}F^{\mu \nu} + \alpha R^{\mu \nu}_{\;\;\;\, \rho \sigma} 
R^{\rho \sigma}_{\;\;\;\, \kappa \lambda}
R^{\kappa \lambda}_{\;\;\;\, \mu \nu} \right]\,,
\end{align}
where $\kappa^2 = 8 \pi G = 1/M_{\rm Pl}^2$, and~$\alpha$ is a coefficient of dimension~${\rm mass}^{-2}$.
One could also include four-derivative operators, such as $R^2$, $R_{\mu\nu}R^{\mu\nu}$ and $R_{\mu\nu\rho\sigma}R^{\mu\nu\rho\sigma}$. However, we are interested in the tidal deformability of neutral black holes in later sections. Since $R$ and $R_{\mu\nu}$ both vanish for a neutral black hole, while~$R_{\mu\nu\rho\sigma}R^{\mu\nu\rho\sigma}$ can be expressed in terms of the first two terms using the Gauss-Bonnet theorem in four dimensions, they do not modify the Schwarzschild geometry. Moreover, one can also verify that they do not contribute to the tidal Love numbers perturbatively for neutral black holes.\footnote{This is due to the essential fact that the linearized equation of motion receives corrections that are either proportional to the background $\{\Bar{R},\, \Bar{R}_{\mu\nu} \}$ or to the linearized GR equation of motion $\{\delta\Bar{R},\, \delta \Bar{R}_{\mu\nu} \}$.} Therefore, $\left(R^{\mu\nu}_{\;\;\;\, \rho \sigma}\right)^3$ is the first non-trivial term that modifies a neutral black hole.\footnote{There is another independent operator involving six derivatives, that is the parity violating term $\tilde{R}^{\mu \nu}_{\;\;\;\,\rho \sigma} R^{\rho \sigma}_{\;\;\;\, \kappa \lambda}
R^{\kappa \lambda}_{\;\;\;\, \mu \nu}$~\cite{Metsaev:1986yb, Endlich:2017tqa}. Such a term would also modify a neutral black hole. However, it mixes parity even and odd perturbations and therefore introduces an ambiguity in the definition of tidal Love numbers~\cite{Cardoso:2018ptl}.}

We consider spherically symmetric, charged black hole solutions of the form 
\begin{align}
    {\rm d}s^2 &= -f_t(r){\rm d}t^2 + \frac{1}{f_r(r)} {\rm d}r^2 + r^2 {\rm d}\Omega^2\,; \nonumber \\
    F &= E(r){\rm d}t\wedge {\rm d}r\,.
\end{align}
The generic solution at $\cal{O}(\alpha)$ is given by,
\begin{align}
\label{metricperturbations}
    f_t(r) &= \frac{(r-r_{-})(r-r_{+})}{r^2}\left[ 1+ \frac{ \alpha \kappa^2   }{ r_{+}^3 ( r- r_{-})}\left\{ -10 \frac{ x^6-1}{x-1} +\frac{6}{7}\frac{r_{-}}{r_{+}} \left( 36 x^6+65\frac{ x^6-x}{x-1}+37 \right) \right. \right. \nonumber \\
     & \hspace{4cm} +\frac{6}{7}\left(\frac{r_{-}}{r_{+}}\right)^2 \left( -62 x^7 -130 x^6-101\frac{ x^6-1}{x-1}\right)\nonumber \\
   & \hspace{4cm}  \left.\left. + \frac{2}{21}\left(\frac{r_{-}}{r_{+}}\right)^3\left( 784 x^8+226 x^7+550 x^6+445\frac{x^6-1}{x-1} \right)  \right\} \right]\,;  \nonumber \\
   f_r(r) & = \frac{(r-r_{-})(r-r_{+})}{r^2}\left[ 1+ \frac{ \alpha \kappa^2   }{ r_{+}^3 ( r- r_{-})}\left\{  98 x^5-10\frac{ x^5-1}{x-1} \right. \right. \nonumber \\
   & \hspace{4cm} +\frac{6}{7} \frac{r_{-}}{r_{+}} \left(-922 x^6+317 x^5 +65\frac{x^5-x}{x-1} +37\right)\nonumber \\
    & \hspace{4cm} +\frac{6}{7}\left(\frac{r_{-}}{r_{+}}\right)^2 \left(2198 x^7-1214 x^6+25 x^5 -101\frac{x^5-1}{x-1}\right) \nonumber \\
   & \hspace{4cm}  \left.\left. + \frac{2}{21}\left(\frac{r_{-}}{r_{+}}\right)^3\left(-12068 x^8+7714 x^7-584 x^6 + 445\frac{x^6-1}{x-1}\right)  \right\} \right] \,; \nonumber \\
    E(r) &= \frac{q}{r^2} \left[1+ 12 \alpha \kappa^2  \left(\frac{1}{r_{+}^4}-\frac{51 r_{-}^2 r_{+}^2}{r^8}+\frac{208 r_{-} r_{+} (r_{-}+r_{+})}{7 r^7}-\frac{9 (r_{-}+r_{+})^2}{2 r^6}\right)  \right],
\end{align}
where $r_{+}$  is the outer horizon, and 
\begin{align}
    x= \frac{r_{+}}{r}\,; \quad \kappa^2 q^2  = 2 r_{-}r_{+}\,.
\end{align}
Note that $r_{+}$ is always the outer horizon, but the inner horizon receives correction from $\alpha$, {\it i.e.}, $r_{-}+ {\cal O}(\alpha)$ is the inner horizon. A neutral black hole corresponds to $r_{-}=0$. In the extremal limit, $r_{+} = r_{-} = r_{h}$, and the above solution reduces to 
\begin{align}
    f_t(r) &= \frac{(r-r_{h})^2}{r^2}\left[ 1+ \frac{ \alpha \kappa^2   }{ r_{h}^4 }\left\{   - \frac{472}{21}\frac{r_h}{r}-\frac{440}{21}\left(\frac{r_h}{r}\right)^2-\frac{136 }{7 }\left(\frac{r_h}{r}\right)^3 -\frac{376}{21}\left(\frac{r_h}{r}\right)^4 \right. \right. \nonumber \\
   & \hspace{4cm}  \left.\left. -\frac{344 }{21 }\left(\frac{r_h}{r}\right)^5 -\frac{104 }{7 }\left(\frac{r_h}{r}\right)^6 -\frac{904 }{21 }\left(\frac{r_h}{r}\right)^7 -\frac{224 }{3 }\left(\frac{r_h}{r}\right)^8   \right\} \right]\,;  \nonumber \\
   f_r(r) & = \frac{(r-r_{h})^2}{r^2}\left[ 1+ \frac{ \alpha \kappa^2   }{ r_{h}^4 }\left\{ - \frac{472}{21}\frac{r_h}{r}-\frac{440}{21}\left(\frac{r_h}{r}\right)^2-\frac{136 }{7 }\left(\frac{r_h}{r}\right)^3 -\frac{376}{21}\left(\frac{r_h}{r}\right)^4 \right. \right. \nonumber \\
   & \hspace{4cm}\left.\left. -\frac{344 }{21 }\left(\frac{r_h}{r}\right)^5 +\frac{2920 }{7 }\left(\frac{r_h}{r}\right)^6 -\frac{4408 }{3 }\left(\frac{r_h}{r}\right)^7 +\frac{3448 }{3 }\left(\frac{r_h}{r}\right)^8  \right\} \right]\,; \nonumber \\
    E(r) &= \frac{q}{r^2} \left[1+ \frac{12 \alpha \kappa^2}{r_h^4}  \left\{1-51\left(\frac{ r_{h}}{r}\right)^8+ \frac{416}{7}\left(\frac{r_h}{ r}\right)^7-18\left(\frac{r_h}{ r}\right)^6\right\}  \right]\,.
\end{align}
Obviously, the charge of the extremal black hole is corrected at ${\cal O}(\alpha)$:
\begin{align}
    Q = \frac{1}{4\pi} \int \star F = \pm q \left( 1+ 12 \frac{\alpha \kappa^2 }{r_h^4} \right) = \pm\frac{\sqrt{2} r_h}{\kappa}\left( 1+ 12 \frac{\alpha \kappa^2 }{r_h^4} \right)\,,
\end{align}
where we have used~$r_\pm = r_h$. There is also an ${\cal O}(\alpha)$ correction to the relation between the ADM mass and the horizon $r_h$, 
\begin{align}
    M =  r_h \left( 1 +\frac{236}{21}\frac{ \alpha  \kappa^2 }{ r_h ^4} \right).
\end{align}
It follows that the charge to mass ratio of an extremal black hole is 
\begin{align}
    \frac{|Q|}{M} = \frac{\sqrt{2}}{\kappa} \left( 1 + \frac{16}{21}\frac{\alpha \kappa^2}{r_h^4}\right)\,. 
\end{align}
Since the weak gravity conjecture demands that $\frac{\kappa|Q|}{\sqrt{2}M} \ge 1 $, we have 
\begin{align}
   \boxed{ \alpha \ge 0}\,.
\end{align}
This is a non-trivial constraint on the effective theory in Eq.~\eqref{eqn:eft}. \footnote{We notice that in a five dimensional theory with a compact dimension, applying the WGC to various black holes implies that the coefficient of  $\left({\cal R}^{AB}_{\quad CD}\right)^3$ (in five dimensions) must be negative \cite{Aalsma:2022knj}. However, it is not clear yet how it may impact the Wilson coefficient of $\left(R^{\mu\nu}_{\;\;\;\, \rho \sigma}\right)^3$ in four dimensions.  }
It must be emphasized that from the EFT point of view, it is natural to include all four-derivative operators composed of $R_{\mu\nu\rho\sigma}$ and $F_{\mu\nu}$. For example, the operators $(F_{\mu \nu}F^{\mu \nu})^2$ and $W_{\mu\nu\rho\sigma}F^{\mu \nu}F^{\rho \sigma}$ contribute to the charge-to-mass ratio of extremal BHs at leading order ${\cal O}(r_h^{-2})$, which is larger than the one resulting from $\left(R^{\mu\nu}_{\;\;\;\, \rho \sigma}\right)^3$. We further comment on this in the conclusions after the discussion on TLNs.

The dominant constraint depends however on the UV theory one considers; in particular, assuming that $(F_{\mu \nu}F^{\mu \nu})^2$ arises by integrating out heavy charged fermions and that $\left(R^{\mu\nu}_{\;\;\;\, \rho \sigma}\right)^3$ is obtained by integrating out a much lighter scalar field, the bound resulting from the latter could be the dominant one if there is a large hierarchy of masses between the two particles. In other words, we consider a slightly tuned EFT in which the leading order contribution comes from $\left(R^{\mu\nu}_{\;\;\;\, \rho \sigma}\right)^3$.
The simple theory we considered above is just for the purpose of demonstrating the connection between the WGC and TLNs. 
A detailed analysis will be carried out in a following work.

Let us now switch to study tidal Love numbers of neutral black holes in the same effective theory, and find the implications of the above constraint on the tidal deformability of black holes.

\section{Tidal Love numbers}
In this Section we review the basics of the computation of tidal Love numbers. We start from their definition in the Newtonian regime, and then move to their relativistic computation for massless spin-1 and spin-2 tidal perturbations in full GR, assuming for simplicity a Schwarzschild black hole background. The interested reader can find a more comprehensive discussion in Refs.~\cite{Hui:2020xxx, Charalambous:2021mea,Charalambous:2022rre}.

\subsection{Newtonian limit}
Tidal Love numbers are defined as the response coefficients of a spherically-symmetric body under the action of external tidal perturbations. 
Consider a spherical body of mass~$M$ placed at the origin of a Cartesian coordinate frame. One can adiabatically apply a static external gravitational field $U_\text{\tiny ext}$ perturbing the body. In the multipole expansion, this field can be written as
\be
U_\text{\tiny ext} =-\sum_{\ell = 2}^\infty \frac{(\ell-2)!}{\ell!}\mathcal{E}_{L} r^L\,,
\ee
in terms of the distance from the origin $r$, the multi-index $L\equiv i_1\cdots i_\ell$, and the symmetric trace-free multipole moments~$\mathcal{E}_{L}$. In response to the external perturbation, the body will deform and develop internal multipole moments given by
\be
I_{L} =\int {\rm d}^3x~\delta\rho(\vec x) x^{\langle L \rangle} \,,
\ee
as a function of the body's mass density perturbation $\delta\rho$ and $x^{\langle L \rangle} \equiv x^{i_1}\cdots x^{i_\ell}$.

Adopting spherical coordinates, the external source and induced response can be expanded in terms of spherical harmonics~$Y^{m}_{\ell}$:
\be
\mathcal{E}_{\ell m}\equiv \mathcal{E}_{L}\int_{\mathbb{S}^2} {\rm d} \Omega~n^L Y^{m\,*}_{\ell}\,;\quad 
I_{\ell m}\equiv I_{L}\int_{\mathbb{S}^2} {\rm d}\Omega~n^L Y^{m\,*}_{\ell}\,,
\ee
where ${\rm d} \Omega \equiv \sin\theta\,{\rm d}\theta\, {\rm d}\phi$, and~$n^i \equiv x^i/|\vec x|$. One can then write the total potential of the system as
\be 
U_\text{\tiny tot}=
-\frac{G M}{r}
-\sum_{\ell=2}^\infty \sum_{m=-\ell}^\ell Y_{\ell m} 
\left[\frac{(\ell-2)!}{\ell !} \mathcal{E}_{\ell m}  r^\ell -
\frac{(2\ell-1)!!}{\ell !} 
\frac{I_{\ell m}}{r^{\ell+1}}
 \right]\,.
\ee
Assuming an adiabatic and weak external tidal perturbation, linear response theory dictates that the response multipoles should be proportional to the perturbing multipole moments as
\be
G I_{\ell m}\left(\omega\right)
=-\frac{\left(\ell-2\right)!}{(2\ell-1)!!}\lambda_{\ell m}(\omega) r_h^{2\ell+1} \mathcal{E}_{\ell m}\left(\omega\right)\,,
\ee
in terms of the characteristic size of the object $r_h$. The dimensionless coefficients~$\lambda_{\ell m}$ describe the response. They are given in terms of the perturbation's frequency~$\omega$ in the external inertial frame as
\be
\label{eq:klm}
	\lambda_{\ell m} \simeq k_{\ell m} + {\rm i}\nu_{\ell m}\left(\omega - m \Omega\right) + \dots\,,
\ee
as a function of the azimuthal harmonic number $m$ and the body's angular velocity $\Omega$. The real term in Eq.~\eqref{eq:klm} describes the conservative response, 
and the corresponding coefficients $k_{\ell m}$ are called tidal Love numbers, while the imaginary contribution ${\rm i}\nu_{\ell m}$ describes dissipation effects.

So far our discussion has been entirely based on the Newtonian regime, which is only a long-distance approximation to the full general relativistic picture. 
In the following we will therefore move to the computation of the tidal Love numbers for a massless spin-1 and spin-2 tidal perturbations, generalising the results found above to a fully relativistic theory. For the purposes of our discussion we will focus on static perturbations $(\omega = 0)$, and we will outline the TLN computation for a Schwarzschild black hole. 

\subsection{GR: vector TLN}
The vector TLN of a black hole is equivalent to its electric polarizability and magnetic susceptibility. It can be studied by considering a massless spin-1 vector field in the background of a Schwarzschild black hole, which is described by the Maxwell action
\be
S = -\int {\rm d}^4 x \sqrt{-g}\frac{1}{4}F_{\mu\nu}^2.
\ee
Because of the rotational invariance of the background, one can expand the gauge potential $A_\mu$ into perturbations and express them in spherical harmonics as
\be
A_\mu = \sum_{\ell, m} \left(
\begin{array}{c}
a_0 Y^m_{\ell} \\
a_r Y^m_{\ell} \\
a^{(L)} \nabla_i Y^m_{\ell}  +a^{(T)} \epsilon_i^{\; j} \nabla_j Y^m_{\ell} 
\end{array}
\right),
\ee
where the index $``i"$ runs over the coordinates on the two sphere, $\nabla_i$ and $\epsilon_i^{\;j}$ are the covariant derivative and  Levi-Civita tensor with respect to $\gamma_{ij}$ on the 2-sphere, respectively. The variables~$a^{(L)}$ and~$a^{(T)}$ denote the longitudinal and transverse perturbations. On the 2-sphere, the variables $a_0$,~$a_r$ and~$a^{(L)}$ are parity-even, while~$a^{(T)}$ is parity odd. Thus they decouple. They are related to each other by electromagnetic duality in four dimensions, and  therefore have equal TLN. For simplicity, we focus on~$a^{(T)}$ for the computation of the TLN.

Introducing the variable $\Psi_\text{\tiny V} = a^{(T)}$ and going to tortoise coordinates ${\rm d} r_\star = {\rm d} r/f$, where $f = 1-r_h/r$, one can write down the action for this perturbation as~\cite{Hui:2020xxx}
\be
S = \int\rd t\rd r_\star \Bigg(\frac{1}{2}\dot\Psi_\text{\tiny V}^2-\frac{1}{2}\left(\frac{\partial\Psi_\text{\tiny V}}{\partial r_\star}\right)^2-\frac{\ell(\ell+1)}{2r^2}f \Psi_\text{\tiny V}^2\Bigg)\,.
\ee
The corresponding equation of motion then takes the form
\be
\left(- \frac{\partial^2}{\partial t^2} +\frac{\partial^2}{\partial r_\star^2} - f(r)\frac{\ell(\ell+1)}{r^2}\right)\Psi_\text{\tiny V} = 0.
\ee
Focusing on the static limit $\omega = 0$ and assuming that the background is asymptotically flat at spatial infinity ($f \to 1$ as $r \to \infty$), one finds that $\Psi_\text{\tiny V}$ can be expanded asymptotically as
\be
\label{vectorTLN}
\Psi_\text{\tiny V} \simeq c_1 r^{\ell+1}\left[1+\cdots+ k_\text{\tiny V}^{(\ell)} \left(\frac{r}{r_h}\right)^{-2\ell-1}+\cdots\right]\,.
\ee
The first term $\sim r^{\ell+1}$ denotes the external tidal field applied at spatial infinity, while the second term $\sim r^{-\ell}$ encodes the quadrupolar response. The coefficient~$k_\text{\tiny V}^{(\ell)}$ denotes the axial (magnetic) vector tidal Love number, and its value depends on the assumed background geometry. 

An obvious problem with the above definition is a possible ambiguity due to an overlap between the source series generated by the gravitational non-linearity and the response contribution~\cite{Kol:2011vg,LeTiec:2020spy,Charalambous:2021mea}, where subleading corrections to the source appear to have the same power in~$r$ as the response in the physical case~$\ell \in \mathbb{N}$. In order to get around this ambiguity and properly define the Love numbers through a matching procedure, one needs to compute the graviton corrections to the source term and subtract them from the full GR solution. An alternative approach consists of an analytic continuation to the unphysical region $\ell \in \mathbb{R}$~\cite{LeTiec:2020spy}, where the source and response series do not overlap. However obtaining such a solution can be challenging, and for simplicity we will ignore it in the rest of the paper.

Furthermore, we stress that the functional form of the field profile can be more complicated than Eq.~\eqref{vectorTLN} due to the presence of logarithmic corrections multiplying the response contribution, coming from EFT loop integrals and resulting into a running Love number for~$\ell \gtrsim 3$. Such logarithms are, however, found to cancel out when summing over all loop diagrams for Schwarzschild black holes~\cite{Ivanov:2022hlo}.

To conclude this discussion, let us comment on the relation between the static response coefficient calculated above and the coefficients which appear in the point-particle effective field theory approach. This connection is important in order to provide a gauge invariant definition of the TLN.

The worldline effective field theory approach is based on the fact that, at very large distance, a black hole behaves as a point particle, and corrections due to the object’s finite size and internal structure are encoded in higher-derivative operators in the effective theory. Focusing on couplings to a tidal magnetic field, the worldline operator can be built starting from the magnetic field $B_{a b} \equiv F_{ab} = \partial_aA_b-\partial_b A_a$, such that the effective field theory action is given by
\begin{align}
S &= \frac{1}{2}\int\rd\tau\left(e^{-1} g_{\mu\nu}\frac{\rd x^\mu}{\rd\tau}\frac{\rd x^\nu}{\rd\tau} - em^2
\right) -\frac{1}{4}\int {\rm d}^4 x F_{\mu\nu}F^{\mu\nu} \nonumber\\
&+  \sum^\infty_{\ell=1} \frac{1}{2\ell!}\int {\rm d} \tau \,e\left[ 
\frac{\lambda_\ell^{(B)}}{2}  \left(  \partial_{( a_1}\cdots \partial_{a_{\ell-1}} B_{a_\ell )_T b} \right)^2  
\right] \,.
\end{align}
The first term describes the free point particle action, expressed in terms of the worldline vielbein~$e$, while the last operator encodes the magnetic susceptibility of the black hole,  proportional to the coefficient $\lambda_\ell^{(B)}$, with $(\cdots)_T$ indicating the trace-free symmetrized part of the enclosed indices. Following Ref.~\cite{Hui:2020xxx}, one can relate this Wilson coefficient with the tidal response coefficient $k_\text{\tiny V}^{(\ell)}$ by matching the gauge-invariant magnetic field as $B_{ij} = \Psi_\text{\tiny V} \nabla_{[i} \epsilon_{j]}^k \nabla_{k} Y_{\ell}^m$, such that
\be
\lambda_\ell^{(B)} =  k_\text{\tiny V}^{(\ell)} (-1)^\ell   \frac{\pi^{\frac{3}{2}}}{2^{\ell-2}} \frac{\Gamma(\frac{1}{2}-\ell)}{ \Gamma(\frac{1}{2})^2} r_h^{2\ell+1}\,.
\ee
This result shows that the coupling to $B^2$ operators in the worldline EFT is absent when the magnetic polarizability vanishes, and that they have the same sign for any $\ell \in \mathbb{N}$.

\subsection{GR: tensor TLN}

The spherical symmetry of the background geometry allows us to decompose metric perturbations into scalar, vector and tensor spherical harmonics: 
\begin{align} \label{eqn:hdecomp}
h_{tt} &= \sum_{\ell,m}  f_t(r)H_0(t,r) Y_\ell^m\,;  \nonumber\\
h_{tr} &= \sum_{\ell,m}  H_1(t,r)Y_\ell^m\,; \nonumber\\
h_{rr} &= \sum_{\ell,m}   f_r(r)^{-1}H_2(t,r)Y_\ell^m\,; \nonumber\\
h_{ti} &= \sum_{\ell,m}\left[{\cal H}_0(t,r)\nabla_i Y_\ell^m +h_0(t,r)\epsilon_i^{\;j}\nabla_j  Y_\ell^m \right]\,; \nonumber\\
h_{ri} &=\sum_{\ell,m}\left[ {\cal H}_1(t,r)\nabla_i Y_\ell^m+ h_1(t,r)\epsilon_i^{\;j}\nabla_j  Y_\ell^m \right]\,; \nonumber\\
h_{ij} &= \sum_{\ell,m}r^2\left[{\cal K}(t,r)\gamma_{ij} Y_\ell^m +G(t,r)\nabla_{(i}\nabla_{j)_T}Y_\ell^m +h_2(t,r) \epsilon_{(i}^{\; \;k}\nabla_{j)}\nabla_{k}Y_\ell^m
\right]\,.
\end{align}
The perturbations~$ H_0, H_1, H_2, {\cal H}_0, {\cal H}_1, {\cal K},G$ belong to the parity-even sector, while $h_0, h_1, h_2$ belong to the parity-odd sector on the 2-sphere. At the linearized level, the two types of perturbations decouple in any theory that is parity invariant (such as~\eqref{eqn:eft}).

For proof of principle, we focus on parity-odd perturbations. Since there is only one physical combination of these degrees of freedom, we will work in Regge--Wheeler gauge~\cite{Regge:1957td}, defined by the condition $h_2 = 0$.
Following Ref.~\cite{DeFelice:2011ka}, one can introduce an additional auxiliary field~$\chi$, such that the action becomes
\be
\begin{aligned}
S_\text{\tiny RW}=\int \rd t \rd r\,\bigg[2\chi\left(\dot h_1+\frac{2}{r}h_0-h_0'\right)-\chi^2+\frac{2f+2rf'-\ell(\ell+1)}{r^2}f\,h_1^2 + \frac{\ell(\ell+1)-2f-2r f'}{r^2 f} h_0^2
\bigg]\,.
\end{aligned}
\ee
We can now integrate out $h_0$ and $h_1$ to obtain an action for $\chi$ only. Their equations of motion then set
\begin{align}
h_0 =-\frac{r f}{(\ell-1)(\ell+2)}\left(2\chi+r\chi'\right)\,; \qquad 
h_1= -\frac{r^2}{(\ell-1)(\ell+2)f}\,\dot \chi\,.
\end{align}
Lastly, introducing the Regge--Wheeler perturbation variable $\Psi_\text{\tiny RW} \equiv \sqrt{\frac{2r^2}{(\ell-1)(\ell+2)}}    \chi$, the action reduces to
\be
S_\text{\tiny RW}=\int\rd t\rd r_\star\Bigg( \frac{1}{2}\dot \Psi_\text{\tiny RW}^2-\frac{1}{2}\left(\frac{\partial\Psi_\text{\tiny RW}}{\partial r_\star}\right)^2-\frac{1}{2}V_\text{\tiny RW}(r)\Psi_\text{\tiny RW}^2
\Bigg)\,; \qquad 
V_\text{\tiny RW}(r) = f\frac{\ell(\ell+1)}{r^2}-\frac{3ff'}{r}\,.
\ee
The equation of motion of $\Psi_\text{\tiny RW}$ takes the form
\be
\bigg(- \frac{\partial^2}{\partial t^2} +\frac{\partial^2}{\partial r_\star^2} - V_\text{\tiny RW}(r) \bigg)\Psi_\text{\tiny RW} = 0\,.
\ee
As before, focusing on the static limit and assuming asymptotic flatness at spatial infinity, one finds that $\Psi_\text{\tiny RW}$ can be expanded asymptotically as
\be
\Psi_\text{\tiny RW} = c_1 r^{\ell+1} \left[1+\cdots+k_\text{\tiny RW}^{(\ell)} \left(\frac{r}{r_h}\right)^{-2\ell-1}+\cdots\right]\,.
\ee
The coefficient $k_\text{\tiny RW}^{(\ell)}$ denotes the axial (magnetic) tensor tidal Love number.

In the worldline effective field theory approach, we consider coupling the black hole point particle to gravity. To determine the tidal response of the point particle and match it to the Love numbers found in the full theory, one has to include higher-derivative worldline couplings to the graviton, usually built from the Weyl tensor $C_{\mu \nu \rho \sigma}$. The magnetic part can then be constructed as $B^{(2)}_{ab|c} \equiv C_{0abc}$, such that the corresponding worldline effective action at quadratic order in the graviton fluctuation $g_{\mu\nu} = \eta_{\mu\nu} + 2h_{\mu\nu}/M_\text{\rm Pl}$ is given by
\begin{align}
S &= \frac{1}{2}\int\rd\tau\left( e^{-1} g_{\mu\nu}\frac{\rd x^\mu}{\rd\tau}\frac{\rd x^\nu}{\rd\tau} - em^2
\right) +\int\rd^4 x \sqrt{-g} \left(\frac{R}{2 \kappa^2} + \dots \right) \nonumber \\
&+\sum_{\ell=1}^\infty\frac{1}{2\ell!} \int\rd \tau \frac{\lambda_\ell^{(C_B)}}{2}\left(\partial_{(a_1}\cdots \partial_{a_{\ell-2}}B^{(2)}_{a_{\ell-1}a_\ell)_T|b}\right)^2.
\end{align}
The last operator encodes the magnetic susceptibility of the black hole. The Wilson coefficient $\lambda_\ell^{(C_B)}$ can be related to the tidal response coefficient $k_\text{\tiny RW}$ using the matching $C_{0rij} \xrightarrow{r\to\infty} 
\nabla_{[i} \epsilon_{j]}^k \nabla_{k} Y_{\ell}^m
r^{-1}\Psi_\text{\tiny RW}$, to obtain~\cite{Hui:2020xxx}
\be
\lambda^{(C_B)}_\ell =  -k_\text{\tiny RW}^{(\ell)}(-1)^\ell \frac{\ell+1}{\ell}   \frac{\pi^{\frac{3}{2}}}{2^{\ell-2}} \frac{\Gamma(\frac{1}{2}-\ell)}{ \Gamma(\frac{1}{2})^2} r_h^{2\ell+1}.
\ee
This equation shows therefore how to relate the response coefficients in the worldline EFT to the tidal Love number computed within GR, and it highlights that they have opposite sign for any~$\ell \in \mathbb{N}$.

\section{Neutral black holes in $\left(R^{\mu\nu}_{\;\;\;\, \rho \sigma}\right)^3$}
In this Section we describe the calculation of the tidal Love numbers for vector and tensor perturbations in the case of neutral black holes, including a $\left(R^{\mu\nu}_{\;\;\;\, \rho \sigma}\right)^3$ correction in the Lagrangian. Recall that the corrected black hole background metric in this case is given by Eq.~\eqref{metricperturbations} with~$r_-= 0$: 
\begin{align}
\label{neutralBHmetric}
f_t &= \frac{(r-r_h)}{r}\left[ 1+ \frac{ \alpha \kappa^2   }{r r_h^3}\left(-\frac{10}{r^5} \frac{r_h^6-r^6}{r_h-r} \right) \right]\,; \nonumber \\
  f_r & = \frac{(r-r_h)}{r}\left[ 1+ \frac{ \alpha \kappa^2  }{ r r_h^3}\left(  98 \left(\frac{r_h}{r} \right)^5-\frac{10}{r^4}\frac{ r_h^5-r^5}{r_h-r} \right) \right]\,.  
\end{align}

\subsection{Vector TLN}
The presence of a term $\left(R^{\mu\nu}_{\;\;\; \rho \sigma}\right)^3$ in the action for a spin-1 tidal field has the effect of modifying the background metric of a neutral black hole, as shown above, but otherwise leaves the vector field equation of motion invariant. The latter is given by
\be
\nabla_\nu F^{\mu \nu} = 0\,,
\ee
in terms of the vector field strength $F^{\mu \nu}$. The corresponding axial perturbation satisfies then the equation of motion in the static limit
\be
\frac{\rd^2\Psi_{\ell}}{\rd r_\star^2} - f_t(r)\frac{\ell(\ell+1)}{r^2} \Psi_{\ell } = 0\,,
\ee
where the tortoise coordinate is now defined as ${\rm d} r_\star = {\rm d} r/ \sqrt{f_r f_t}$.
This equation can be solved perturbatively in the coupling strength $\alpha$. Considering the multipole moments $\ell = 1$ and $\ell=2$, and imposing regularity of the solution at the black hole horizon $r_h$, the solution at zeroth order in $\alpha$ has the following form 
\begin{align}\label{psizeroth}
\Psi_{\ell = 1}^{(0)} &= c_1 r^2\,; \nonumber \\
\Psi_{\ell = 2}^{(0)} &= c_1 \left(r^3 - \frac{3}{4} r^2 r_h \right)\,,
\end{align}
in terms of a dimensional constant $c_1$.
We stress that we have made use of the fact that the metric perturbations $f_r$ and $f_t$ coincide at zeroth-order in $\alpha$. The absence of an induced dipolar or quadrupolar term implies that the dipole and axial vector TLNs of an unperturbed neutral black hole are both zero.

At first order in $\alpha$, the metric perturbations $f_r$ and $f_t$ receive a correction, as shown in Eq.~\eqref{neutralBHmetric}. Expanding the solution as $\Psi = \Psi^{(0)} + \alpha \Psi^{(1)}$, one finds that the first-order solutions are given by 
\begin{align}
\Psi_{\ell = 1}^{(1)} &=
- c_1 \frac{\kappa^2}{r_h^2}\left[4 \left( \frac{r_h}{r} \right) + 3\left( \frac{r_h}{r} \right)^2 + \frac{12}{5}\left( \frac{r_h}{r} \right)^3-22\left( \frac{r_h}{r} \right)^4 \right]\,; \nonumber \\
\Psi_{\ell = 2}^{(1)} &= -c_1 \frac{\kappa^2}{r_h} \left[ \frac{15}{2} \left( \frac{r}{r_h} \right)^2 + \frac{219}{20} \left( \frac{r_h}{r} \right)^2 - \frac{197}{5} \left( \frac{r_h}{r} \right)^3 + \frac{33}{2} \left( \frac{r_h}{r} \right)^4 \right]\,.
\end{align}
Combined with~\eqref{psizeroth}, the full solutions in the asymptotic regime $r \to \infty$ thus take the form
\begin{align}
\Psi_{\ell = 1} &= \Psi_{\ell = 1}^{(0)} + \alpha \Psi_{\ell = 1}^{(1)} \simeq c_1 r^2 \left[1 + k_\text{\tiny V}^{(\ell =1)} \left(\frac{r_h}{r} \right)^3 \right]\,;\nonumber \\
\Psi_{\ell = 2} &= \Psi_{\ell = 2}^{(0)} + \alpha \Psi_{\ell = 2}^{(1)} \simeq c_1 r^3 \left[1 + k_\text{\tiny V}^{(\ell =2)} \left(\frac{r_h}{r} \right)^5 \right]\,.
\end{align}
From these one can extract the tidal response coefficients as
\begin{align}
k_\text{\tiny V}^{(\ell =1)} &= - 4 \frac{\alpha \kappa^2}{r_h^4}\,; \nonumber \\
k_\text{\tiny V}^{(\ell =2)} &= - \frac{219}{20} \frac{\alpha \kappa^2}{r_h^4}\,.
\end{align}
The corresponding Wilson coefficients in the worldline effective field theory approach are then
\begin{align}
\lambda_{\ell = 1}^{(B)} &= - 16 \pi \frac{\alpha \kappa^2}{r_h}\,; \nonumber \\
\lambda_{\ell = 2}^{(B)} &= -\frac{73}{5} \pi r_h \alpha \kappa^2\,.
\end{align}
This result shows that a positive value for the coupling $\alpha$, as dictated by the WGC, would result in a negative axial vector tidal Love number.

Proceeding to the computation for higher $\ell$, we know that the extraction of the Love number is not well defined, as discussed in the previous Section. In particular, the perturbative solution at~$\ell=3$ is given by 
\begin{align}
     \Psi_{\ell =3}& = c_1 \left( r^4-\frac{4 r^3 r_h}{3}+\frac{2 r^2 r_h^2}{5} \right) 
     + c_1 \alpha \kappa^2 \Bigg\{ -\frac{40}{3} \left(\frac{r}{r_h}\right)^3 + 8  \left(\frac{r_h}{r}\right)^2  \nonumber \\
     &\qquad  -  \left[ \frac{184616}{1225}  - \frac{704}{7} \ln \left(\frac{r}{r_h}\right) \right] \left(\frac{r_h}{r}\right)^3 + {\cal O}\left[\left(\frac{r_h}{r}\right)^4, \left(\frac{r_h}{r}\right)^4 \ln \left(\frac{r}{r_h}\right)  \right] \Bigg\}\,.
\end{align}
By identifying the coefficient of~$r^{-\ell}$, one finds
\begin{align}
    ``k_\text{\tiny V}^{\ell =3}" = \frac{\alpha \kappa^2}{r_h^4}\left[-\frac{184616}{1225}  + \frac{704}{7} \ln \left(\frac{r}{r_h}\right) \right]\,,
\end{align}
where a running behaviour appears.

Let us also stress that, in principle, a non-vanishing vector TLN may arise if one considers four derivative term containing the field strength~$F_{\mu \nu}$, {\it e.g.}, $R^{\mu \nu \rho \sigma} F_{\mu \nu} F_{\rho \sigma}$ studied in~\cite{Garcia-Saenz:2022wsl}. However, since this term neither modify the background solution nor generate any linear tensor response for neutral black holes, we have neglected it in our analysis.

\subsection{Tensor TLN}
In this Section we compute the TLN in the parity-odd sector of tensor perturbations, where only $h_0$, $h_1$ and $h_2$ in Eq.~\eqref{eqn:hdecomp} are relevant. We further choose Regge-Wheeler gauge $h_2=0$. The quadratic action, in terms of $h_0$, $h_1$ in a neutral black hole background, of the theory \eqref{eqn:eft} is
\begin{align} \label{eqn:actionh0h1}
    S_{(2)} &= \int {\rm d}t {\rm d}r \bigg[\frac{1}{4\kappa^2}\left( h_0'  - \frac{2}{r} h_0 + \dot{h}_1 \right)^2 + \frac{j^2 +2 - 2 f - 6r f' - r^2 f''}{4 \kappa^2 r^2 f } h_0^2 + f \frac{4f + 6rf' + r^2 f''}{4 \kappa^2 r^2} h_1^2 \nonumber \\ 
   & \hspace{2cm} + \alpha \tilde{{\cal L}}[h_0,\;h_1] \bigg]\,,
\end{align}
where~$j^2 = \ell(\ell+1)$, and~$f=1- r_h/r$. We have written $f_t = f + \alpha \delta f_t$, $f_r=f + \alpha \delta f_r$ and collected all~${\cal O}(\alpha)$ terms in $ \alpha \tilde{{\cal L}}$. The explicit form of the latter is given in Appendix~\ref{app:explicitform}. The first line in~\eqref{eqn:actionh0h1} is the same as in GR, while the second line includes terms coming from the ${\cal O}(\alpha)$ corrections to the background metric as well as the quadratic action from the $\alpha \left(R^{\mu\nu}_{\;\;\;\, \rho \sigma}\right)^3$ operator. 

We now extract the gauge-invariant variable by introducing once again an auxiliary field~$\chi$ as in \cite{DeFelice:2011ka},
\begin{align}
 &   S_{(2)}[h_0,\; h_1,\; \chi] \nonumber \\
  &= \int \rd t \rd r \bigg[\frac{1}{4\kappa^2}\left( h_0'  - \frac{2}{r} h_0 + \dot{h}_1 \right)^2 + \frac{j^2 +2 - 2 f - 6r f' - r^2 f''}{4 \kappa^2 r^2 f } h_0^2 + f \frac{4f + 6rf' + r^2 f''}{4 \kappa^2 r^2} h_1^2 \nonumber \\ 
   & \hspace{2cm}   - \frac{1}{4\kappa^2} \left(  h_0'  - \frac{2}{r} h_0 + \dot{h}_1 - \chi\right)^2 + \alpha \tilde{{\cal L}}[h_0,\;h_1]\bigg] .
\end{align}
It is obvious that $\chi$ is a gauge invariant variable~\cite{Martel:2005ir}, since its equation of motion gives
\begin{align}
    \chi =   h_0'  - \frac{2}{r} h_0 + \dot{h}_1 \,.
\end{align}
Since we are treating $\alpha$ perturbatively, one can introduce 
\begin{align}
    h_0 &= h_0^{(0)} + \alpha h_0^{(1)}\,; \nonumber \\
    h_1 &= h_1^{(0)} + \alpha h_1^{(1)}\,; \nonumber \\
    \chi &= \chi^{(0)} + \alpha \chi^{(1)}\,.
\end{align}
 Keeping terms up to first order in~$\alpha$, the quadratic action becomes 
 \begin{align} \label{eqn:pertaction}
     &   S_{(2)}[h_0^{(0)},\; h_1^{(0)},\; \chi^{(0)},\;h_0^{(1)},\; h_1^{(1)},\; \chi^{(1)}] \nonumber \\
     & = S_{(2)}[h_0^{(0)},\; h_1^{(0)},\; \chi^{(0)}]  + \alpha \int \rd t \rd r \bigg[ \frac{1}{2\kappa^2}\left( h_0^{(0)}{}'  - \frac{2}{r} h_0^{(0)} + \dot{h}_1^{(0)} \right)\left( h_0^{(1)}{}'  - \frac{2}{r} h_0^{(1)} + \dot{h}_1^{(1)} \right) \nonumber \\     
    & \hspace{1cm}  + \frac{j^2 +2 - 2 f - 6r f' - r^2 f''}{2 \kappa^2 r^2 f } h_0^{(0)}h_0^{(1)} + f \frac{4f + 6rf' + r^2 f''}{2 \kappa^2 r^2} h_1^{(0)}h_1^{(1)} \nonumber \\ 
    &   \hspace{1cm}  - \frac{1}{2\kappa^2} \left(  h_0^{(0)}{}'  - \frac{2}{r} h_0^{(0)} + \dot{h}_1^{(0)} - \chi^{(0)}\right) \left(  h_0^{(1)}{}'  - \frac{2}{r} h_0^{(1)} + \dot{h}_1^{(1)} - \chi^{(1)}\right)  \bigg]\,.
 \end{align}
It is clear that~$h_0^{(1)}$ and~$h_1^{(1)}$ are now Lagrange multipliers. Their equations of motion give
\begin{align}
    h_0^{(0)} & = \frac{r_h-r}{j^2-2} \left( 2\chi^{(0)} +  r \chi^{(0)}{}'\right)\, ; \nonumber\\
    h_1^{(0)} & = \frac{r^3 }{(j^2-2)(r_h-r)}  \dot{\chi}^{(0)}\,.
\end{align}
After substituting these back to the action \eqref{eqn:pertaction}, we are left with an action for $\chi^{(0)}$ and $\chi^{(1)}$ only.\footnote{Of course, even though~$h_0^{(1)}, h_1^{(1)}$ are gone in the Lagrangian, there are still equations of motion relating them to~$\chi^{(0)}$, $\chi^{(1)}$.} The equation of motion for $\chi$ from this action can be recast in the Regge-Wheeler form after rescaling $\chi$ by 
\begin{align}
    \chi^{(0)} +\alpha  \chi^{(1)} = \frac{\sqrt{2(j^2 -2)}}{r } \left( \Psi^{(0)} +\alpha  \Psi^{(1)} \right). 
\end{align}
The resulting equations are
\begin{align} \label{eqn:Psieom}
   \left( -\partial^2_t + \partial^2_{r^*} + \left(1-\frac{r_h}{r}\right)\frac{3 r_h - j^2 r}{r^3} \right) \Psi^{(0)} &=0\,; \nonumber \\
    \left( -\partial^2_t + \partial^2_{r^*} + \left(1-\frac{r_h}{r}\right)\frac{3 r_h - j^2 r}{r^3} \right) \Psi^{(1)} &= {\cal S}[\Psi^{(0)}]\,,
\end{align}
where $\partial_{r^*} = \left(1-r_h/r\right) \partial_r$. The explicit form of the  source term~${\cal S}[\Psi^{(0)}]$ is given in Appendix~\ref{app:explicitform}. 

\begin{figure}[t]
    \centering
    \includegraphics[scale=0.285]{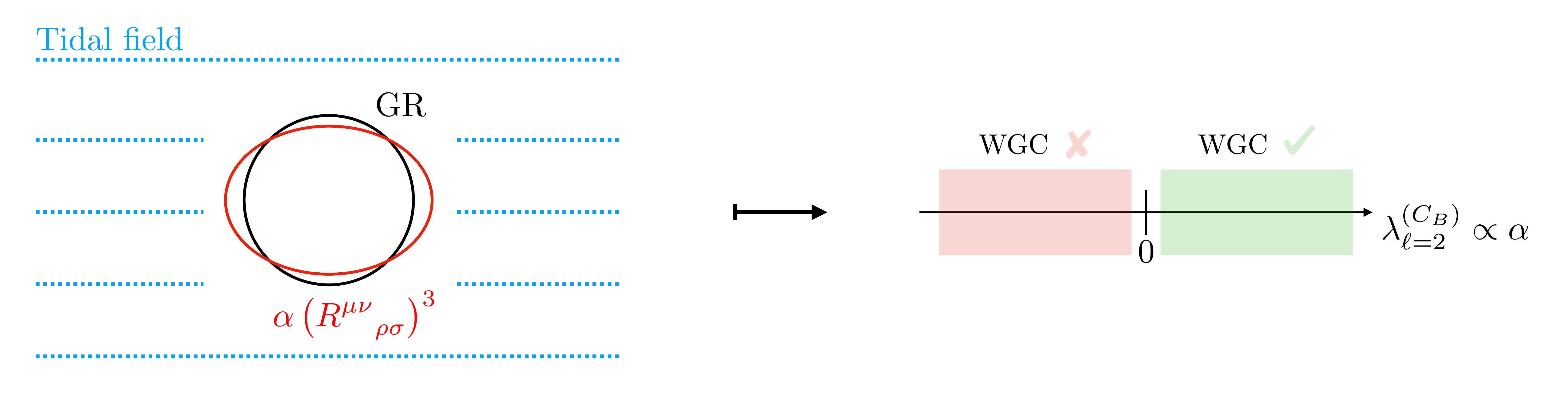}
    \caption{Pictorial representation of the tidal deformability of a black hole with a $\left(R^{\mu\nu}_{\;\;\;\, \rho \sigma}\right)^3$ term and corresponding sign of the tensor tidal Love number for the multipole $\ell = 2 $.}
    \label{fig:TLN}
\end{figure}

We are now ready to study the static limit where $\Psi$ only depends on $r$. It should be emphasized that the static limit should be taken in the gauge invariant combination $\Psi$ instead of $h_0$ and $h_1$. Then the solution for $h_0$, $h_1$ are calculated from $\Psi$. Following the procedure in the previous sections, the regular solution for $\Psi_{\ell=2}$ is 
\begin{align}
    \Psi_{\ell =2} = c_1 r^3 + \frac{\alpha \kappa^2}{r_h} c_1 \left( - 240 \left(\frac{r_h}{r}\right)^2 + 250 \left(\frac{r_h}{r}\right)^3 \right).
\end{align}
Therefore the tidal Love number in the Regge-Wheeler variable for $\ell =2$ is\footnote{A similar computation has been performed in Ref.~\cite{Cai:2019npx} for the metric tensor perturbations $(h,H)$, without however estimating the TLN for the more appropriate gauge invariant variable. } 
\begin{align}
    k_\text{\tiny RW}^{(\ell =2)} = -240 \frac{\alpha \kappa^2}{r_h^4}.
\end{align}
Correspondingly, the coupling in the worldline effective action is 
\begin{align}
   \boxed{ \lambda^{(C_B)}_{\ell = 2} = 480 \pi r_h \alpha \kappa^2}\,.
\end{align}
This result shows that a positive value for the coupling $\alpha$, as dictated by the WGC, would result in a positive parity-odd tidal Love number. See Fig.~\ref{fig:TLN} for a pictorial representation.

When one continues the analysis to higher $\ell$, the story becomes subtle, as mentioned earlier, since the identification of the Love number is unclear. For instance, the perturbative solution at~$\ell=3$ is 
\begin{align}
\Psi_{\ell =3}& = c_1 r^4 + c_1 \alpha \kappa^2 \Bigg\{ -\frac{25}{3} \left(\frac{r}{r_h}\right)^3 - 480  \left(\frac{r_h}{r}\right)^2  \nonumber \\
     &\qquad + \frac{125}{49} \left[- 391  +448 \ln \left(\frac{r}{r_h}\right) \right] \left(\frac{r_h}{r}\right)^3   + {\cal O}\left[\left(\frac{r_h}{r}\right)^4, \left(\frac{r_h}{r}\right)^4 \ln \left(\frac{r}{r_h}\right)  \right] \Bigg\}\,.
\end{align}
One can still identify the coefficient of $1/r^{\ell}$, which is 
\begin{align}
    ``k_\text{\tiny RW}^{(\ell =3)}" = \frac{125}{49}  \frac{\alpha \kappa^2}{r_h^4} \left[- 391  +448 \ln \left(\frac{r}{r_h}\right) \right]
\end{align}
However, it is unclear whether this is still the Love number as it is no longer the dominating tail. The same issue appears when one consider a tidal scalar field in the same background~\cite{Ivanov:2022hlo}. 

We conclude this section with a final remark on the assumed response function. In particular, we have computed only the ``linear'' response of the black hole to the external tidal perturbation. In principle a $\alpha\left(R^{\mu\nu}_{\;\;\;\, \rho \sigma}\right)^3$ term would generate also a nonlinear response when the non-linear source term in the equation of motion is taken into account. At each order of the perturbative series in~$\alpha$, the non-linear source are simply the product of lower-order terms in the series. For instance, at~${\cal O}(\alpha)$ and $\ell=4$, a source proportional to $\left(C^{\ell =2} \right)^2$ should appear, where $C^{\ell=2}$ is the magnitude of the source tidal field. However, it mixes parity-odd and -even modes, and requires the use of Clebsch-Gordan coefficients between vector and tensor spherical harmonics.  The computation of this response would be quite challenging, and the extraction of the corresponding TLN is still not well defined. We have therefore neglected this contribution for the sake of our discussion.

\section{Conclusions}
Within the low energy effective field theory, extremal black holes should be able to decay according to the WGC. The latter sets non-trivial bounds on the sign of the coefficients of higher-order derivative corrections to GR. On the other hand, the presence of these operators can have an impact on the multipolar structure of black holes.
Indeed, even though  neutral black holes in pure GR are characterized by a vanishing tidal Love number, which measures the static deformability of compact objects under external tidal perturbations, the presence of these  higher-order derivative corrections can modify this result and give rise to a nonzero tidal Love number.

As a proof of concept, in order to explicitly show the connection between TLNs and the WGC, we have focused on a  $\alpha\left(R^{\mu\nu}_{\;\;\;\, \rho \sigma}\right)^3$ term in the effective action of gravity, which can arise in a theory with higher derivatives or by integrating out heavy massive fields~\cite{Goon:2019faz}.
We showed that the WGC dictates only positive values for the coupling $\alpha$, assuming that the coefficients of the four-derivative operators are small enough compared to $\alpha$. By computing the tidal deformability of neutral black holes, which is insensitive to four-derivative operators and receives leading-order correction from $\alpha$, we concluded that only positive values for the tidal Love numbers (in the point particle effective action) are allowed in this specific example. As a consequence, if negative TLNs is measured for neutral BHs, in order to obey the WGC, it signals that there must exist relevant four-derivative operators that correct the charge-to-mass ratio for extremal BHs.

Given the dimensionful nature of this coupling, we stress that integrating out massive fields with a lower cut-off scale would eventually enhance the tidal Love numbers, and that a positive value for the coupling may imply the existence of a very light bosonic field in the theory. On the other hand, the detection of a negative tidal Love number may give bound the coefficients of four derivative operators in order to satisfy the WGC.  Among the four-derivative operators, those composed of $R_{\mu\nu\rho\sigma}$ and $F_{\mu\nu}$, such as $\alpha_1\kappa^4(F_{\mu \nu}F^{\mu \nu})^2$ and $\alpha_3 \kappa^2 W_{\mu\nu\rho\sigma}F^{\mu \nu}F^{\rho \sigma}$, will contribute to the charge-to-mass ratio of extremal BHs, on top of $\alpha\left(R^{\mu\nu}_{\;\;\;\, \rho \sigma}\right)^3$, as  
\begin{align}
 \frac{|Q|}{M} = \frac{\sqrt{2}}{\kappa}\left( 1 + \frac{2 \kappa ^2}{5r_h^2} ( 4\alpha_1 - \alpha_3)  + \frac{16}{21}\frac{\alpha \kappa^2}{r_h^4} \right).
\end{align}
If the TLNs of neutral BHs are negative ($\alpha<0$), the WGC implies that 
\begin{align}
     4\alpha_1 - \alpha_3  >- \frac{40}{21}\frac{\alpha }{r_h^2}.
\end{align}
Also, in this case we expect that the TLNs of extremal BHs would be proportional to some combination of the operators' Wilson coefficients as $k_2 \propto {\cal O}(\alpha_1,\alpha_3) \kappa^2/r_h^2$, showing that the connection between TLNs and WGC would hold as well for those objects. We leave a detailed analysis of these operators to future work~\cite{WGCextremal}.

Constraints on the sign and size of the couplings of higher-derivative terms in the effective action of gravity can also be set from other observables, such as scattering amplitudes. For example, the presence of terms like Gauss-Bonnet operators are known to affect also the graviton 3-point interaction, leading to violations of causality unless the value of the coupling $\alpha$ is very small, independently of its sign~\cite{Camanho:2014apa}. Similarly, higher-derivative terms may impact the graviton four-point function, which can be used, using arguments of causality and unitarity, to impose constraints on their coupling coefficients~\cite{Metsaev:1986yb,Gruzinov:2006ie,Caron-Huot:2021rmr,deRham:2021bll,Caron-Huot:2022ugt}.
Furthermore, we stress that higher-derivative operators like $\left(R^{\mu\nu}_{\;\;\;\, \rho \sigma}\right)^3$ may impact as well on the black hole entropy, which would be corrected by a term proportional to $\alpha$, with the entropy increasing only for positive values of $\alpha$~\cite{Cai:2019npx}.

Finally, a complete unitarity analysis of all derivative-six operators, composed of $R^{\mu\nu}_{\;\;\;\,\rho \sigma}$ and $F_{\mu\nu}$, along the line of \cite{Hamada:2018dde,Bellazzini:2019xts,Arkani-Hamed:2021ajd}, would be valuable.

\subsubsection*{Acknowledgments}
We thank L. Aalsma, A. Kehagias, D. Lombardo, T. Noumi, R. Penco, A. Riotto, I. Rothstein, L. Santoni, C. Vafa and M. Wiesner for interesting comments and discussions.
V.DL. is supported by funds provided by the Center for Particle Cosmology at the University of Pennsylvania. 
The work of J.K. and S.W. is supported in part by the DOE (HEP) Award DE-SC0013528.

\pagebreak

\appendix
\section{Explicit form of the action and equation of motion for $\Psi$} \label{app:explicitform}
For completeness, we report the explicit form of the action $\tilde{\cal L}[h_0, h_1]$ in Eq.~\eqref{eqn:actionh0h1} and the source term for the equation of motion of the perturbation field $\Psi$ in the tensor sector of Eq.~\eqref{eqn:Psieom}. The former is given by 
\begin{align}
& \tilde{\cal L}[h_0, h_1] = \nonumber \\
 &\frac{\left(j^2-2\right) r \left(5 r^4 r_h+5 r^3 r_h^2+5 r^2 r_h^3+161 r r_h^4-226 r_h^5+5 r^5\right)-6 \left(j^2-2\right)^2 r^2 r_h^4+42 \left(r-r_h\right) \left(8 r-15 r_h\right) r_h^4 }{2 r^8 \left(r-r_h\right) r_h^3} h_0^2 \nonumber \\
 &+\frac{3 r_h \left(39 r_h-10 j^2 r\right) \dot{h_0}' h_0}{r^7}+\frac{3 r_h \left(4 \left(j^2-6\right) r+27 r_h\right) h''_0 h_0}{2 r^6}+\frac{6 r_h \left(3 r-4 r_h\right) h'''_0 h_0}{r^5}\nonumber \\
 &+\frac{3 r_h \left(r_h-r\right) h''''_0 h_0}{r^4}+\frac{6 \left(j^2 r-r_h\right) r_h \ddot{h_0} h_0}{r^4 \left(r-r_h\right){}^2}+\frac{\left(6 r_h^2-9 r r_h\right) \ddot{h_0}' h_0}{r^3 \left(r-r_h\right){}^2}+\frac{3 r_h \ddot{h_0}'' h_0}{r^2 \left(r-r_h\right)}\nonumber \\
 &+\frac{3 r_h \left(\left(13 j^2+10\right) r^2-3 \left(5 j^2+24\right) r_h r+66 r_h^2\right) \dot{h_1} h_0}{r^7 \left(r-r_h\right)}+\frac{3 r_h \left(\left(4-8 j^2\right) r+9 r_h\right) \dot{h_1}' h_0}{2 r^6}\nonumber \\
 &+\frac{6 r_h \left(3 r_h-2 r\right) \dot{h_1}'' h_0}{r^5}+\frac{3 \left(r-r_h\right) r_h \dot{h_1}''' h_0}{r^4}+\frac{3 r_h \dddot{h_1} h_0}{r^2 \left(r-r_h\right){}^2}+\frac{3 r_h \dddot{h_1}' h_0}{r^2 \left(r_h-r\right)}\nonumber \\
 &+\frac{h_1^2 \left(j^2-2\right) \left(5 r^6+6 \left(j^2-2\right) r_h^4 r^2-3 \left(2 j^2+17\right) r_h^5 r+58 r_h^6\right)}{2 r^9 r_h^3}\nonumber \\
 &+\frac{3 r_h \left(7 j^2 r^2-5 \left(j^2+5\right) r_h r+21 r_h^2\right) \dot{h_0} h_1}{r^7 \left(r-r_h\right)}+\frac{3  r_h \left(\left(12-8 j^2\right) r+9 r_h\right) \dot{h_0}' h_1}{2 r^6}\nonumber \\
 &+\frac{3  r_h \left(6 r_h-5 r\right) \dot{h_0}''h_1}{r^5}+\frac{3  \left(r-r_h\right) r_h \dot{h_0}'''h_1}{r^4}+\frac{6  r_h \dddot{h_0}h_1 }{r^3 \left(r-r_h\right)}+\frac{3 r_h \dddot{h_0}' h_1 }{r^2 \left(r_h-r\right)}\nonumber \\
 &+\frac{6  \left(j^2-2\right) \left(r-r_h\right){}^2 r_h h_1'' h_1}{r^7}+\frac{3 r_h \left(4 \left(j^2+4\right) r-33 r_h\right) \ddot{h_1} h_1 }{2 r^6}+\frac{3  r_h \left(3 r-4 r_h\right) \ddot{h_1}'h_1}{r^5}\nonumber \\
 &+\frac{3 r_h \left(r_h-r\right) \ddot{h_1}'' h_1}{r^4}+\frac{3  r_h \ddddot{h_1}h_1}{r^2 \left(r-r_h\right)}-\frac{6 \left(j^2-2\right) \left(5 r-7 r_h\right) \left(r-r_h\right) r_h h_1' h_1}{r^8}.
\end{align}
We stress that in principle the expression can be further simplified. The source term is instead given by 
\begin{align}
   & \frac{1}{\kappa^2}{\cal S}[\Psi^{(0)}] =  \nonumber \\
 &  \left(
 \frac{4 (r-r_h)^2 \left(5 r^5+5 r^4 r_h+5 r^3 r_h^2+5 r^2 r_h^3-211 r r_h^4+194 r_h^5\right)}{r^8 r_h^3}
 \right)  \partial^2_r\Psi^{(0)} \nonumber \\
 & -\left(
 \frac{2 (r-r_h) \left(5 r^6-5 r^5 r_h-5 r^4 r_h^2-5 r^3 r_h^3-5 r^2 r_h^4-203 r r_h^5+212 r_h^6\right)}{r^9 r_h^3}
 \right) \partial_r\Psi^{(0)} \nonumber \\
 & - \Bigg(
\frac{2 j^2 \left(5 r^6-432 r^2 r_h^4+900 r r_h^5-473 r_h^6\right)}{r^9 r_h^3}+\frac{6 \left(5 r^7-10 r^6 r_h+432 r^2 r_h^5-915 r r_h^6+488 r_h^7\right)}{r^{10} r_h^3}
 \Bigg)\Psi^{(0)}\,,
\end{align}
where we have repeatedly used the equation of motion~\eqref{eqn:Psieom} for $\Psi^{(0)}$  to reduce the higher-order derivative terms from $\left(R^{\mu\nu}_{\;\;\;\, \rho \sigma}\right)^3$. 

\bibliographystyle{JHEP}
\bibliography{WGCTLN.bib}
\end{document}